# Synergistic Photon Management and Strain-Induced Band Gap Engineering of Two-Dimensional MoS$_2$ Using Semimetal Composite Nanostructures

Xiaoxue Gao[1,*], Sidan Fu[1,*], Tao Fang[1], Xiaobai Yu[1], Haozhe Wang[2], Qingqing Ji[2,3], Jing Kong,[2,**] Xiaoxin Wang,[1,**] Jifeng Liu[1,**]

[1] Thayer school of Engineering, Dartmouth College, 15 Thayer Drive, Hanover, New Hampshire 03755, USA

[2] Department of Electrical Engineering and Computer Science, Massachusetts Institute of Technology, 77 Massachusetts Avenue, Cambridge, Massachusetts 02139, USA

[3] Current address: School of Physical Science and Technology, ShanghaiTech University, Shanghai 201210, China

[*] Authors contribute equally

[**]Corresponding authors: Jifeng.Liu@dartmouth.edu; Xiaoxin.Wang@dartmouth.edu; JingKong@mit.edu

## Abstract

2D MoS$_2$ attracts increasing attention for its application in flexible electronics and photonic devices. For 2D material optoelectronic devices, light absorption of the molecularly thin 2D absorber would be one of the key limiting factors in device efficiency, and conventional photon management techniques are not necessarily compatible with them. In this paper, we show two semimetal composite nanostructures for synergistic photon management and strain-induced band gap engineering of 2D MoS$_2$: (1) pseudo-periodic Sn nanodots, (2) conductive SnO$_x$ (x<1) core-shell nanoneedle structures. Without sophisticated nanolithography, both nanostructures are self-assembled from physical vapor deposition. 2D MoS$_2$ achieves up to >15x enhancement in absorption at λ=650-950 nm under Sn nanodots, and 20-30x at λ=700-900 nm under SnOx (x<1) nanoneedles, both spanning from visible to near infrared regime. Enhanced absorption in MoS$_2$ results from strong near field enhancement and reduced MoS$_2$ band gap due to the tensile strain induced by the Sn nanostructures, as confirmed by Raman and photoluminescence spectroscopy. Especially, we demonstrate that up to 3.5% biaxial tensile strain is introduced to 2D MoS$_2$ using conductive nanoneedle-structured SnOx (x<1), which reduces the band gap by ~0.35 eV to further enhance light absorption at longer wavelengths. To the best of our knowledge, this is the first demonstration of a synergistic *triple-functional* photon management, stressor, and conductive electrode layer on 2D MoS$_2$. Such synergistic photon management and band gap engineering approach for extended spectral response can be further applied to other 2D materials for future 2D photonic devices.

**Key Words**

# 1. Introduction

Recent advances in 2D material fabrication techniques enables further investigation into ultrathin 2D materials' properties. $MoS_2$, a prototypical member of transition metal dichalcogenide family, has an indirect bandgap around 1.2 eV in bulk form. When thinned down to monolayer, it would transform into a direct bandgap of about 1.9 eV [1]. Monolayer $MoS_2$ based field-effect transistors also have been reported for their extraordinary on-off current ratio and mobility [2]. These superior optical and electronic properties demonstrate promising applications of 2D $MoS_2$ in novel optoelectronic device in the visible range.

However, the optical absorption of 2D $MoS_2$ is too low to serve as efficient photodetector or photovoltaic devices. Conventional photon management strategies for 2D materials such as Fabry–Pérot microcavity [3] or guided resonance in photonic crystal [4] would need complicated fabrication processes and the enhanced light absorption only covers a relatively narrow spectral range. On the other hand, while multi-layer $MoS_2$ has been applied to achieve broadband photoresponse from 450 to 2700 nm [5], the quantum efficiency is still limited to ~10% in the visible regime (at λ~500 nm) and <5% in the near infrared (NIR) regime at λ>800 nm.

Strain engineering, i.e. tuning the electronic and photonic properties by introducing strain into the material [6], creates new possibilities for light absorption enhancement of 2D materials. Several approaches to introduce strain into 2D material have been reported. Since 2D material has to be supported by a substrate, most of the work focus on deforming or pre-patterning the bulk substrate to transfer strain into the 2D material [7]. These methods either have specific requirements for substrates such as flexibility, or need specially designed mechanical system to apply strain which are not compatible with optoelectronic device integration. Other methods, such as laser-induced local heating [7] or stacking heterostructures of two different 2D material [8], have also also investigated. However, the induced strain is hard to control. Specific substrates such as Pt are also needed for strain-engineering via heterostructures of 2D materials [8], and the strain can be lost upon transfer to other substrates (e.g. Si or silica) for practical device fabrication. Stressor layers [9], commonly used in traditional semiconductor industry to improve carrier mobility, are thin film layers designed to transfer strain to the active electronic or photonic materials. Peña et al. [10] have recently reported the fabrication of several electrically insulating dielectric thin film stressor capping layers on 2D $MoS_2$. On the other hand, conductive stressors would be very useful for 2D optoelectronic devices to serve as an electrode simultaneously.

In this work, we introduce semimetal composite nanostructures that not only effectively improve the light absorption of the 2D $MoS_2$ via photon management, but also simultaneously extends spectral absorption in 2D $MoS_2$ through strain-induced band gap engineering. Two-dimensional $MoS_2$ achieves up to 15x enhancement in absorption at λ=650-950 nm under self-assembled pseudo-periodic Sn nanodots, and 20-30x at λ=700-900 nm under self-assembled $SnO_x$ (x<1) nanoneedles, both spanning from visible to near infrared regime. Our study shows that Sn composite nanostructures take advantage of ultrahigh permittivity of Sn [11-13], which greatly improve the light absorption of $MoS_2$ underneath via local field enhancement. The physical vapor deposition (PVD) of these Sn-based self-assembled nanostructures is also fully compatible with 2D materials in maintaining their perfect lattice. In this way, Sn serves as a representative of semimetal material system beyond plasmonic metals and high index dielectrics for enhancing optical interactions with 2D materials. Furthermore, the extended spectral absorption of 2D $MoS_2$ at longer wavelengths >700 nm is synergistically enabled by strain-induced band gap shrinkage in conjunction with the photon management. Our self-assembled conductive $SnO_x$ (x<1) core-shell nanoneedle structure introduces up to 3.5% biaxial tensile strain to the 2D $MoS_2$ and decreases the band gap by ~0.35 eV. At the same time, it also serves as a conductive electrode. To the best of our knowledge, this is the first demonstration of a facile and synergistic triple-functional photon management, stressor, and conductive electrode layer on 2D $MoS_2$. Such synergistic photon management and band gap engineering approach for extended spectral response can be further applied to other 2D materials for future 2D optoelectronic devices.

## 2. Results and Discussion

### 2.1 The Sn Nanodots/MoS$_2$/Quartz System

The photon management effect of Sn nanodot is first explored. We will later examine the synergistic photon management effect and strain-induced band gap engineering of 2D $MoS_2$ by $SnO_x$ (x<1) nanoneedles. In our experiment, chemical vapor deposition (CVD)-prepared pristine 2D $MoS_2$ is transferred to a fused quartz substrate, on which nominally 12 nm thick Sn is thermally evaporated at 0.15 Å/s. Note that Sn dewetts the surface of the fused quartz substrate, hence the "12-nm thickness" based on the deposition rate is nominal. Pseudo-periodic Sn nanodots are self-assembled from the thermal evaporation process and their morphology could be controlled through the deposition rate [11-12]. Details about Sn nanodot deposition have been reported in previous work [11-12].

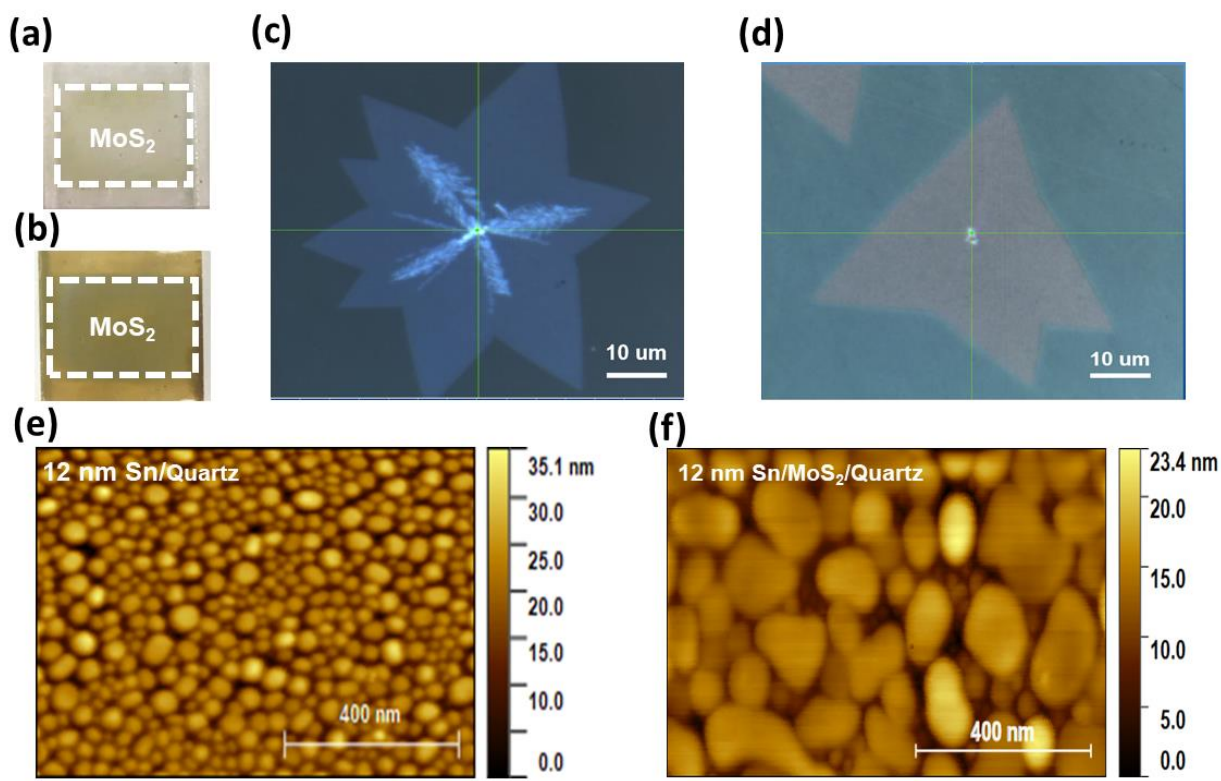

**Figure 1.** Photos of MoS$_2$/quartz samples: (a) before and (b) after nominally 12 nm Sn nanodot deposition. The region with MoS$_2$ is indicated by the dashed rectangle. (c) and (d) show the corresponding optical microscopy photos. The area coverage of MoS$_2$ flakes is approximately 22.5%. The brighter regime is due to double or multiple layer overlapping of MoS$_2$. The color change before and after Sn nanodot deposition is due to the change in reflectance spectra. (e) AFM image of nominal 12 nm Sn nanodots on quartz, i.e. without MoS$_2$. (f) AFM image of nominal 12 nm Sn nanodots deposited on MoS$_2$/quartz, corresponding to (b) and (d).

**2.1.1 Optical Microscopy and Surface Morphology: Figures 1a** and **1b** compare the photos of MoS$_2$/quartz sample before and after the Sn nanodot deposition. Before Sn nanodot deposition, the region with MoS$_2$ is almost transparent under white light illumination. After Sn nanodot fabrication, the region with MoS$_2$ shows a clear color contrast compared to its surrounding region without MoS$_2$ (**Figure 1b**). Corresponding optical microscopy images in **Figures 1c** and **1d** compare the MoS$_2$ flakes before and after the Sn nanodot fabrication, showing clear differences in optical color contrast due to the modifications in absorption/reflectance spectra, as will be detailed later. The MoS$_2$ area coverage on the quartz samples is ~22.5% from statistical analyses of the optical microscopy images. The brighter regions in the graph is due to overlapping of bilayer and multiple layer MoS$_2$. We also use atomic force microscopy (AFM) to characterize the Sn nanodot morphology on quartz vs. that on MoS$_2$. **Figures 1e** and **1f** demonstrate the morphology of Sn nanodots on the regions without and with MoS$_2$ on the same sample, respectively. Sn nanodots directly deposited on the quartz substrate (i.e. regions without MoS$_2$) are much smaller than those on MoS$_2$. Statistically, the size of Sn nanodot is 32.7 ± 11.5 nm on quartz substrate vs. 100.6 ± 37.4 nm on MoS$_2$. The air gap between Sn nanodots is 16.7 ± 6.9 nm on quartz substrate vs. 26 ± 16.5 nm on MoS$_2$. The difference in surface energy between quartz and MoS$_2$ is the main cause of such variation. These narrow air gaps are important for the photon management. Due to the boundary condition of Maxwell's equation, the optical power at the air gap is much stronger than that at Sn nanodot since the dielectric constant of air is much smaller than that of Sn. Consequently, the interaction between light and MoS$_2$ right under the airgap regions is significantly enhanced.

**2.1.2 MoS$_2$ Optical Absorption Enhancement**: To quantitatively measure this optical absorption enhancement, UV-Vis-NIR spectrophotometer equipped with an integrating sphere is used to measure the transmittance (T) and reflectance (R) spectra. The absorption spectrum is therefore calculated by:

$$Absorption\ (Abs) = 1 - T - R \qquad (1)$$

A complication, though, is that the morphology of Sn nanodot on quartz differs from that on MoS$_2$, as we demonstrated above by AFM images. Therefore, directly comparing the absorption of Sn/MoS$_2$/quartz with its surroundings (without MoS$_2$) could induce some error in evaluating the enhanced absorption in MoS$_2$. Thus, we simulate the overall Sn nanodot/MoS$_2$/quartz absorption spectrum as well as the individual contributions from Sn nanodots and MoS$_2$ in COMSOL Multiphysics software using the experimentally measured Sn nanodot morphology shown in **Figure 1f**. The refractive index n and extinction coefficient k of Sn nanodots have been reported in previous work [11], while those of MoS$_2$ are derived from the transmittance and reflectance spectra of the pristine MoS$_2$/quartz sample used in this study by transfer matrix method (See Supporting Information, Section 1). We also use previously demonstrated and experimentally verified pseudo-periodic approximation to model the optical spectra [11,12]. **Figure 2a** shows that the simulated overall absorption spectrum of Sn nanodot/MoS$_2$/quartz agrees very well with the experimental results, especially at 650-1200 nm. This agreement validates our model to further separate the contribution of MoS$_2$ from that of Sn nanodots, as detailed in **Figure 2a**. At shorter wavelength <700 nm, Sn nanodot contributes more to the overall absorption while MoS$_2$ weighs more at longer wavelength.

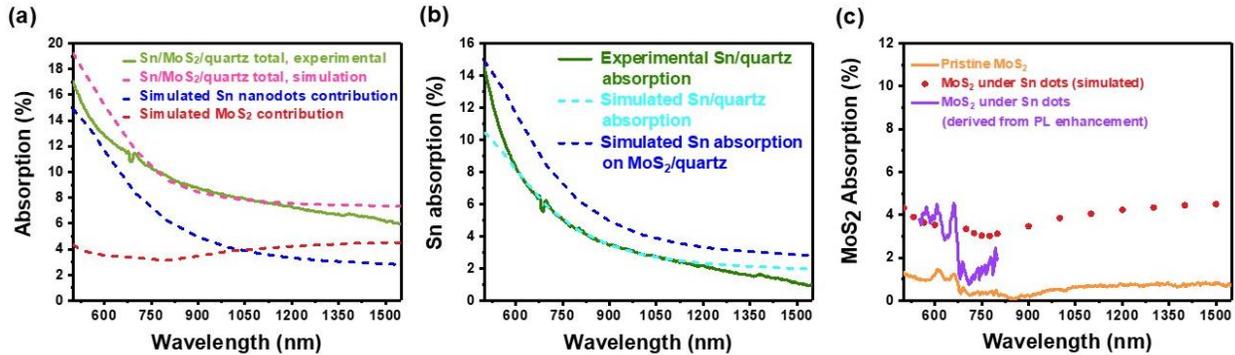

**Figure 2.** (a) Experimentally measured Sn nanodot/MoS$_2$/quartz absorption spectrum (solid light green line) in comparison with COMSOL simulation (dashed magenta line), showing very good agreement with each other especially at 650-1200 nm. The simulated contributions of Sn nanodot absorption and MoS$_2$ absorption spectra are also shown by the dashed blue line and the dashed red line, respectively, which add up to the dashed magenta line. (b) Comparison of COMSOL simulated (dashed cyan line) and experimentally measured (solid green line) Sn nanodot/quartz absorption spectra. The simulated Sn nanodot absorption spectrum on MoS$_2$ is also shown (dashed blue line). The morphological difference between Sn nanodots on MoS$_2$ (Figure 1f) vs. those on quartz (Figure 1e) mainly induces a ~1.4x increase in absorption, corresponding to the same increase in areal coverage of Sn nanodots, while the shapes of the absorption spectra are similar. (c) Absorption spectrum of 2D MoS$_2$ in the Sn nanodot/MoS$_2$/quartz structure derived from the simulation in (a) (red dots), compared with that derived from PL enhancement shown in Figure 3a (purple line), The absorption spectrum of pristine MoS$_2$/quartz is also shown as a reference. A small amount of absorption is observed in the infrared regime from the reference sample due to a small fraction of bilayer and multilayer MoS$_2$, which is also enhanced by Sn nanodots photon management and leads to notable infrared absorption of MoS$_2$ in the Sn nanodot/MoS$_2$/quartz structure.

To further verify our model for Sn nanodots, **Figure 2b** compares the modeled and experimentally measured absorption spectra of the Sn nanodots on quartz (with diameter=32.7 ± 11.5 nm and gap=16.7 ± 6.9 nm), which also agrees very well with each at 600-1200 nm. This agreement again confirms the reliability of our model. Furthermore, compared to the absorption of Sn nanodot on quartz, we found that the modeled Sn nanodot absorption on $MoS_2$ (diameter=100.6 ± 37.4 nm, gap=26 ± 16.5 nm) is 1.4x larger almost irrespective of the wavelength. Since their geometric shapes are similar and sizes are much smaller than the wavelengths of interest (see **Figures 1e** and **1f**), the absorption of Sn nanodot is largely proportional to the areal coverage (i.e. the fraction of area covered by Sn dots). For Sn nanodot on quartz the areal coverage is 40% while for Sn nanodot on $MoS_2$, it is 57%. The areal coverage ratio between Sn nanodot on $MoS_2$ and that on quartz is ~1.4, which is indeed consistent with the optical absorption ratio between them. These analyses further validate the separation of $MoS_2$ absorption from that of Sn nanodots shown in **Figure 2a**.

**Figure 2c** then compares the absorption of 2D $MoS_2$ under Sn nanodots (red dots), as derived from the consistent experimental and theoretical data in **Figure 2a**, with that of pristine $MoS_2$/quartz reference sample. We observe absorption at $\lambda>1000$ nm for the pristine $MoS_2$/quartz reference sample due to a small amount of bilayer and multiplayer $MoS_2$ (e.g. brighter parts in **Figure 1c**), which agrees well with previously reported experimental studies [5, 14] showing that multilayer $MoS_2$ could exhibit absorption beyond the band gap of monolayer $MoS_2$. With photon management, 2D $MoS_2$ shows up to 15x enhancement in absorption at $\lambda=650-950$ nm under pseudo periodic Sn nanodots. The small absorption of pristine $MoS_2$ in the NIR regime due to double/multilayer regions is also dramatically enhanced by photon management effect after Sn nanodot deposition.

In addition to photon management, another possible factor contributing to the enhanced NIR absorption is the formation of Schottky barrier between the Sn nanostructure and MoS$_2$. As shown in previous work experimentally, the work function of Sn nanodots is around 4.7 eV [15] while the electron affinity of MoS$_2$ is reported to be around 3.92 eV in literature [16]. Therefore, the potential barrier for photoelectron injection from the Fermi level of Sn to the conduction band of MoS$_2$ is only ~0.8 eV. Even light at 1500 nm would have enough energy to excite electrons from Sn to the conduction band of MoS$_2$, which can also substantially enhance the IR absorption of MoS$_2$ under photon management.

**2.1.3 Further Verification by Photoluminescence (PL) and Raman Enhancement:** We further utilize PL enhancement in **Figure 3a** to confirm the absorption enhancement shown in **Figure 2c**, considering that both result from field enhancement in 2D MoS$_2$. In the PL analysis, the incident excitation wavelength is $\lambda_{ex}$=532 nm while the PL spectra span from 550 to 800 nm. Significant PL enhancement is observed in a broad wavelength range from 550 to 800 nm compared to pristine MoS$_2$, consistent with the broad-spectral photon management discussed in **Figure 2**. The relationship between PL enhancement at wavelength $\lambda_{PL}$ and the corresponding electrical field enhancement can be approximated as:

$$\frac{I_{PL}(Sn/MoS_2/quartz)}{I_{PL}(MoS_2)} \approx \left(\frac{|E|^2}{|E_0|^2}\bigg|_{\lambda-ex} \cdot T_{ex}\right) \cdot \left(\frac{|E|^2}{|E_0|^2}\bigg|_{\lambda-PL} \cdot T_{PL}\right) \qquad (2)$$

Here $I_{PL}$ represents the PL intensity at wavelength $\lambda_{PL}$; $E_0$ and E are the electric field in MoS$_2$ before and after Sn nanodot deposition, respectively, evaluated at the excitation wavelength $\lambda_{ex}$=532 nm and the PL wavelength $\lambda_{PL}$. $T_{ex}$ is the transmittance of the excitation laser through the Sn nanodots in order to excite MoS$_2$, and it is approximated by the UV-Vis-NIR spectrophotometer measured transmittance at 532 nm. $T_{PL}$ is the transmittance of the PL signal through the Sn/MoS$_2$ structure in order to be collected by the PL photodetection system, and it is approximated by the transmittance spectrum in the PL spectrum regime of λ=550-800 nm (see Supporting Information Section 2, Figure S1). Therefore, the terms in the first parenthesis on the right-hand side of Equation (2) refer to the enhanced absorption at the excitation wavelength, while those in the second parenthesis refer to the enhanced PL emission at wavelength $\lambda_{PL}$

In Equation (3), the field enhancement at 532 nm, $\frac{|E|^2}{|E_0|^2}|_{\lambda-ex}$, is derived from the integrated Raman peak intensity enhancement shown in **Figure 3b** [11-12]:

$$\frac{I_{E2g+A1g}(Sn\ naodot/MoS_2/quartz)}{I_{E2g+A1g}(MoS_2)} \approx \frac{|E|^4}{|E_0|^4}|_{\lambda-ex} \cdot T_{Raman} \qquad (3)$$

Here, $I_{E2g+A1g}$ represents the integrated Raman peak intensity of $MoS_2$ including both in-plane $E_{2g}$ and out-of-plane $A_{2g}$ modes, as shown in **Figure 3b**. $T_{Raman}$ is the transmittance of the Raman-scattered photons through the Sn nanodot/$MoS_2$ interface, as approximated by the transmittance spectrum shown in Figure S2. Therefore, using data in **Figure 3** and **Equations (2) and (3)**, we derive the corresponding wavelength-dependent electric field and absorption enhancement in $MoS_2$ under Sn nanodot photon management, as shown by the purple line in **Figure 2c**. The results agree well with the COMSOL simulation of $MoS_2$ absorption within ~0.5% difference at λ=550-660 nm. At 680-800 nm, the $MoS_2$ absorption derived from these two methods still show a similar trend, i.e. reaching a minimum around 750 nm before increasing again with the increase of wavelength. The quantitative discrepancy in this spectral regime is due to the very weak absorption of the 2D $MoS_2$ reference sample at 700-800 nm (<<1%), which tends to induce large relative errors in measuring the absorption (A) and deriving the extinction coefficient k in this spectral regime for COMSOL modeling. Overall, the PL enhancement provides another strong evidence to support the broad-band optical absorption enhancement in MoS$_2$ due to ultrahigh refractive index semimetal Sn nanodot photon management.

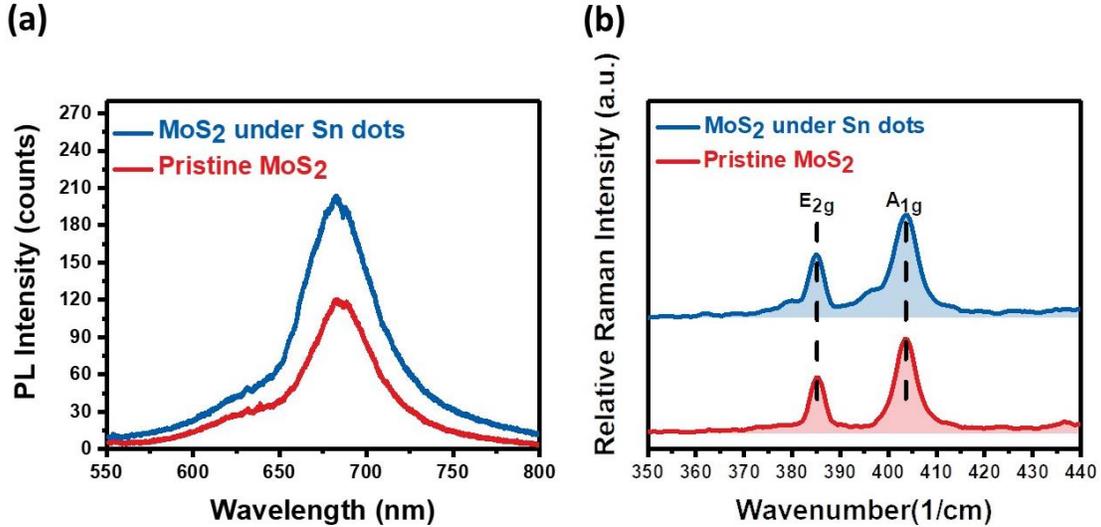

**Figure 3.** (a) PL spectroscopy of MoS$_2$ on Sn/MoS$_2$/quartz sample (blue line) in comparison to that on MoS$_2$/quartz sample (red line), under the excitation wavelength of 532 nm. (b) Raman spectroscopy of 2D MoS$_2$ under 532 nm laser on Sn/MoS$_2$/quartz sample (blue line) in comparison with that on MoS$_2$/quartz sample (red line).

**2.1.4 Strain Analyses:** We also use the Raman spectra in **Figure 3b** to confirm the integrity of the 2D MoS$_2$ lattice and to evaluate the local strain after Sn nanodot deposition. As mentioned earlier, two strongest Raman peaks from MoS$_2$ are observed: in-plane vibration E$_{2g}$ at 384 cm$^{-1}$ and out-of-plane vibration A$_{1g}$ at 408 cm$^{-1}$ [17]. Additionally, little shoulders around E$_{2g}$ and A$_{1g}$ are observed for the Sn nanodot/MoS$_2$/quartz sample. This is due to the inhomogeneous tensile strain around the periphery of the Sn nanodots that causes the Raman peak signals to redshift compared to other regions, as has also been reported in the case of Ag nanodots on MoS$_2$ [18]. The field enhancement at the narrow gaps between Sn nanodot, as mentioned earlier, also enhances the Raman signals from these tensile strained regions. This observation suggests that Sn nanostructure can be applied to strain-engineer MoS$_2$ without affecting the 2D lattice integrity, which will be discussed in more detail in the next section for Sn-rich SnO$_x$ (x<1) nanoneedles on MoS$_2$.

### 2.2. The SnO$_x$ (x<1) Nanoneedle/MoS$_2$/Quartz System

As discussed earlier, the tensile strain in MoS$_2$ induced by Sn nanodots can redshift its bandgap for extended absorption in NIR regime, yet it is not uniform as shown by the separation of the shoulders from the main peaks in **Figure 3b**. To further explore photon management in 2D MoS$_2$ that simultaneously offers strain-induced band engineering, the other semimetal composite nanostructure, Sn-rich SnO$_x$ (x<1) nanoneedle, is fabricated by co-sputtering of Sn and SnO$_2$ [13] onto the MoS$_2$/quartz sample. The sample is then annealed at ~225 °C in nitrogen atmosphere to form the nanoneedle structures based on our previous work [13, 19-20].

**2.2.1 Overview of Microstructures, Optical, and Electrical Properties of SnOx Nanodeedles:** As discussed in Ref. 13 and further detailed in **Figure 4**, Sn-rich SnO$_x$ (x<1) nanoneedles comprise in a Sn core/SnO shell structure. The top-view (backscattered electron mode, BSE) and cross-sectional scanning electron microscopy (SEM) images, together with the corresponding energy-dispersive X-ray spectroscopy (EDS) mapping in **Figure 4a, c,** clearly identify nanoscale Sn rich regions surrounded by tin oxide. The X-ray diffraction (XRD) data in **Figure 4b** further identify two phases: Sn and SnO, which correspond to the Sn cores and SnO cladding observed in **Figure 4a, c**. This is further characterized via high resolution transmission electron microscopy (HRTEM) images in the Supporting Information **Figure S2**, showing rectangular and trapezoidal cross-sections of the nanoneedle structures comprising Sn/SnO core/shells. Similar to the case of Sn nanodots, the nanoscale gaps between ultrahigh refractive index Sn cores (**Figure 4** and **Figure S2**) as well as the high refractive index Sn/SnO nanoneedles (see the dark gap regions in **Figure 4a**) provide effective field enhancement to the 2D MoS$_2$ region underneath the nano-gaps. Indeed, the effective refractive index of these nanoneedle structures are derived to be n~2-3 and k~0.3-1 over visible and NIR light regime, whose trend is similar to that of Sn semimetal (see Supporting Information Section 4 and **Figure S3**). Simultaneously, the nanoneedle network offers high electrical conductivity on the order of 1500-2000 S/cm [19, 20], as has been demonstrated in synergistically enhancing NIR absorption and forming p-n heterojunctions with Ge photodiodes [19]. Therefore, the SnOx nanoneedles can potentially form a conductive electrode on MoS$_2$ and simultaneously enhance its optical absorption.

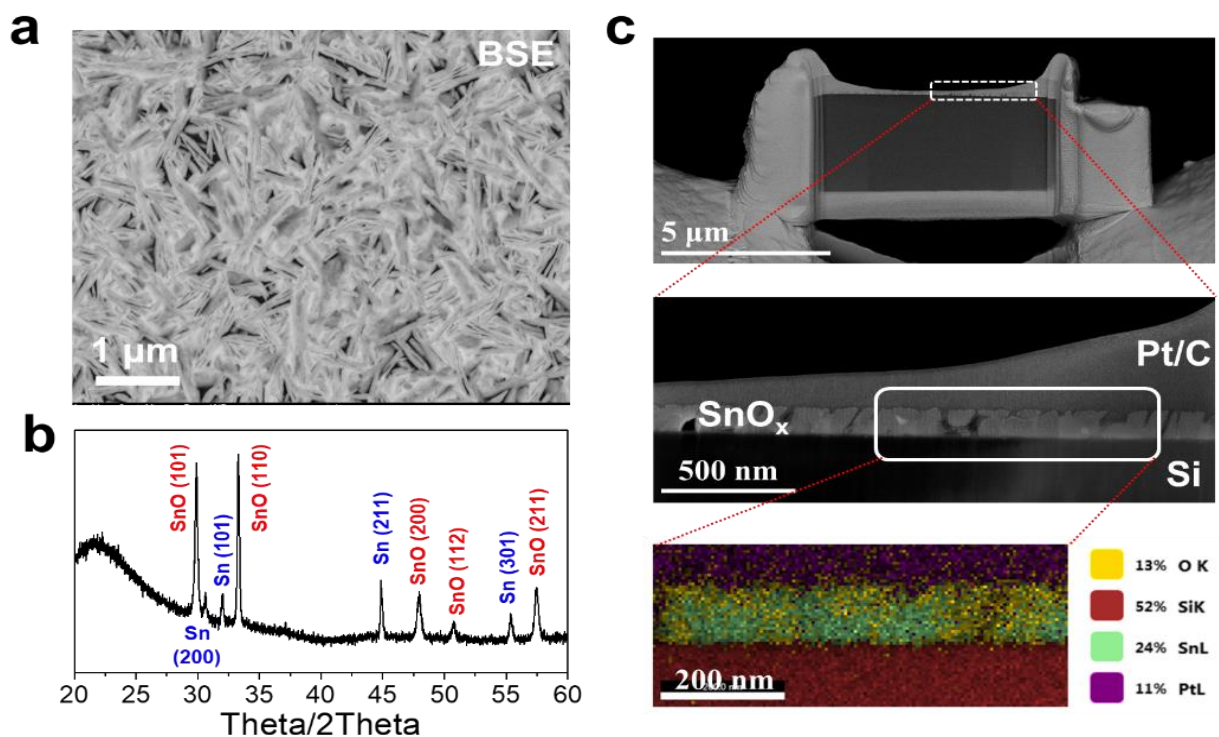

**Figure 4 (a)** Backscattered electron (BSE) mode scanning electron microscopy (SEM) characterization of 115 nm-thick SnO$_x$ nanoneedle-structured thin film from top-down view. Brighter parts indicate Sn cores due to higher atomic weight that increases the yield of backscattered electrons. **(b)** θ-2θ X-ray diffraction (XRD) analysis of SnO$_x$ nanoneedle-structured thin film on fused quartz substrate. Both Sn and SnO phases are observed. **(c) Upper panel** is a cross-sectional view of 115 nm-thick SnO$_x$ nanoneedle-structured thin film on Si substrate, prepared and characterized by an SEM-focused ion beam (FIB) dual beam system. **Middle panel** is a zoom-in of the upper panel showing rectangular and trapezoidal cross-sections of the nanoneedles. **Lower panel** is an energy-dispersive X-ray spectroscopy (EDS) mapping of the SnO$_x$ nanoneedle-structured thin film cross-section clearly showing Sn cores surrounded by tin oxide.

**2.2.2 Optical Absorption Enhancement of MoS₂:** After the SnOx nanodeedle deposition, the SnO$_x$ nanoneedle/MoS$_2$/quartz sample is further characterized by UV-Vis-NIR and Raman spectroscopy. The photo in **Figure 5a** shows that the region with MoS$_2$ is notably darker than the surrounding region without MoS$_2$. As previously reported, the morphology of SnO$_x$ nanoneedles does not vary with substrate material, including 2D materials such as graphene [13]. Therefore, we can directly utilize the optical absorption contrast between regions with and without MoS$_2$ on the same SnO$_x$/MoS$_2$/quartz sample (as shown in **Figure 5a**) to evaluate the MoS$_2$ absorption under SnO$_x$ nanoneedle photon management. 2D MoS$_2$ absorption spectra contrast under 115 nm and 174 nm thick SnO$_x$ (x<1) nanoneedle thin films are shown in **Figure 5b**. Quantitatively, under 115 nm-thick SnO$_x$ nanoneedles, 2D MoS$_2$ peaks at 7% at λ~720 nm. Under 174 nm-thick SnO$_x$ nanoneedles, the absorption of 2D MoS$_2$ reaches 11% at λ ~900 nm. The enhanced absorption covers both visible and near-infrared light region, which explains the darker color contrast in the region with MoS$_2$ compared to its surrounding part without MoS$_2$ on the same SnO$_x$/MoS$_2$/quartz sample, as shown in **Figure 5a**. We also note that, compared to the primary direct gap absorption peak of the pristine MoS$_2$ at λ~660 nm (~1.9 eV), the absorption peaks shift by ~0.2 eV and ~0.5 eV for the 115 and 174 nm SnO$_x$/MoS$_2$/quartz samples, respectively. As will be detailed later, this is primarily due to the strain-induced bandgap shrinkage from the SnO$_x$ stressor layer.

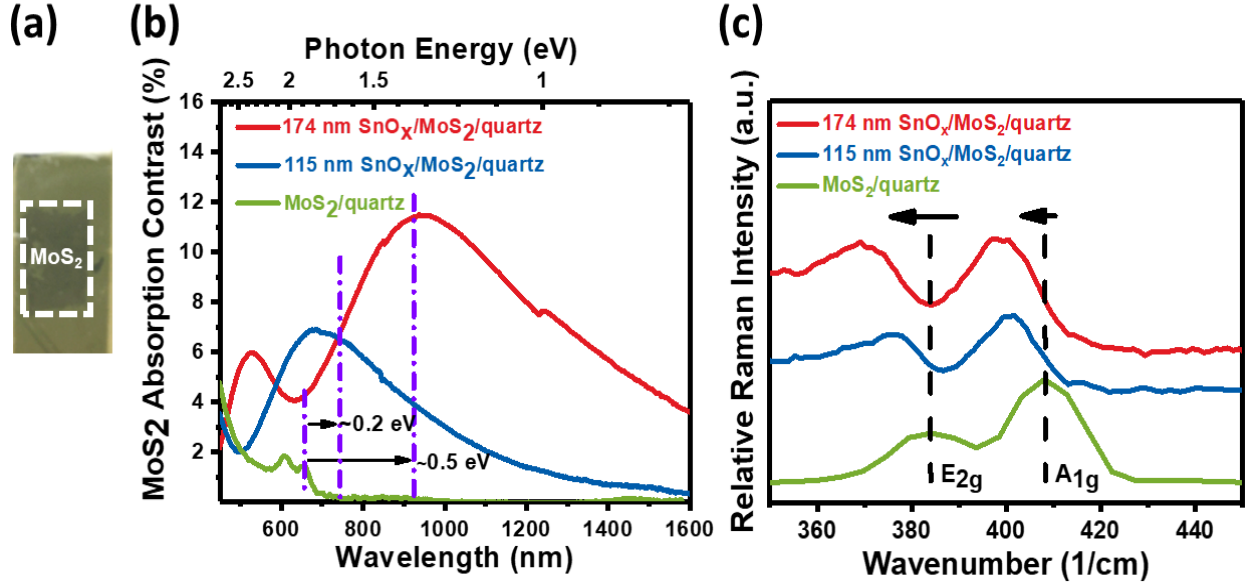

**Figure 5. (a)** A photo of the $SnO_x/MoS_2$/quartz sample, where the region with 2D $MoS_2$ is indicated by the dashed rectangle. **(b)** Absorption spectra of 2D $MoS_2$ in 115 nm $SnO_x/MoS_2$/quartz sample (blue line) and 174 nm $SnO_x/MoS_2$/quartz sample (red line) in comparison with that of pristine $MoS_2$/quartz sample (green line). **(c)** Raman spectra of 2D $MoS_2$ in 115 nm $SnO_x$ /$MoS_2$/quartz sample (blue line) and 174 nm $SnO_x/MoS_2$/quartz sample (red line) in comparison with that of pristine $MoS_2$/quartz sample (green line). For 115 nm and 174 nm $SnO_x$ (x<1) nanoneedle samples, the field enhancement analysis is demonstrated in **Table 1**. The redshift of Raman peak positions from $MoS_2$/quartz sample to $SnO_x/MoS_2$/quartz samples indicates tensile strain introduced into 2D $MoS_2$.

**2.2.3 Further Verification via Raman Enhancement:** As discussed earlier, PL enhancement would be an effective approach to further verify the enhanced absorption in $MoS_2$. However, we found that Sn-rich $SnO_x$ itself also has significant PL at 550-800 nm, most likely due to defect states in SnO, which overlaps that of $MoS_2$ and makes it hard to accurately evaluate the PL enhancement in $MoS_2$ alone. Therefore, we utilize Raman intensity enhancement instead to evaluate the field enhancement at the Raman excitation wavelength as an alternative confirmation approach. In **Figure 5c,** Raman spectra of a pristine $MoS_2$/quartz sample, a 115 nm $SnO_x$ /$MoS_2$/quartz sample, and a 174 nm $SnO_x/MoS_2$/quartz sample are demonstrated, respectively. Characteristic in-plane vibration mode $E_{2g}$ at 384 cm$^{-1}$ and out-of-plane vibration $A_{1g}$ at 408 cm$^{-1}$

[17] are observed. Similar to Equation (3), the Raman peak enhancement is related to the field enhancement by

$$\frac{I_{E2g+A1g}(SnOx/MoS_2/quartz)}{I_{E2g+A1g}(MoS_2)} \approx \frac{|E|^4}{|E_0|^4} \cdot T_{Raman} \qquad (4)$$

Here, $I_{E2g+A1g}$ represents the integrated Raman peak intensity of MoS$_2$, as shown in **Figure 5c**. E and E$_0$ are the electric field in MoS$_2$ with and without SnO$_x$ nanoneedle photon management, respectively. T$_{Raman}$ is the transmittance of the Raman-scattered photons through the SnO$_x$/MoS$_2$ interface.

**Tables 1** lists the detailed MoS$_2$ field enhancement of these two samples derived from Raman peak enhancement. At the excitation wavelength of 514 nm, ~1.7x optical field enhancement is observed in the 115 nm SnO$_x$/MoS$_2$/quartz sample, and ~3.3x in the 174 nm SnO$_x$/MoS$_2$/quartz sample. 2D MoS$_2$ absorption derived from Raman integrated intensity enhancement is also compared with the data measured from UV-Vis-NIR spectrophotometer (**Figure 5b**) in the table, showing very good agreement with each other. Therefore, the Raman data further confirm the field enhancement induced by the SnO$_x$ nanoneedle structure.

**Table 1.** Summary of field enhancement ($|E|^2/|E_0|^2$) and strain in MoS$_2$ derived from Raman peak shift and intensity enhancement for **115 nm** SnO$_x$/MoS$_2$/quartz and **174 nm** SnO$_x$/MoS$_2$/quartz samples compared to pristine MoS$_2$ at the excitation wavelength of 514 nm. The field enhancement estimated from the enhanced Raman scattering confirms the photon management by SnO$_x$ (x<1) nanoneedles, which also agrees well with the absorption enhancement at 514 nm shown in **Figure 5b**. The theoretical strain-induced bandgap shrinkage also largely agrees with the absorption peak shift in **Figure 5b**.

| Thickness of SnO$_x$ film (nm) | Measured MoS$_2$ Raman Peak Intensity Ratio $\frac{I_{E+A}(MoS_2 \ under \ SnOx)}{I_{E+A}(MoS_2)}$ | Transmittance of Raman-Scattered Photons T$_{Raman}$ | MoS$_2$ Field Enhancement $\frac{E^2(MoS_2 \ under \ SnOx)}{E_0^2(MoS_2)}$ | MoS$_2$ Absorption derived from Raman enhancement (%) | MoS$_2$ Absorption Measured by UV-Vis-NIR spectroscopy (%) | Measured Strain in MoS$_2$ (%) | Expected strain-induced Bandgap shrinkage (eV) |
|---|---|---|---|---|---|---|---|
| 115 | 0.67 | 0.23 | 1.7 | 2.4 | 2.2 | 1.7 | 0.17 |
| 174 | 0.90 | 0.085 | 3.3 | 4.6 | 5.3 | 3.5 | 0.35 |

**2.1.4. Strain Analyses**: Furthermore, an important and unique feature beyond photon management effect is that $SnO_x$ nanoneedles also modulate the band gap of 2D $MoS_2$ by introducing uniform biaxial tensile strain, thereby simultaneously achieving triple functions: photon management, band engineering, and conductive electrode on $MoS_2$. In **Figure 5c**, a large redshift of both $E_{2g}$ and $A_{1g}$ Raman peak is observed after $SnO_x$ deposition, together with an increase in the separation between these two peaks. Since the $MoS_2$/quartz sample for nanoneedle deposition is composed of both monolayer and a few layer 2D $MoS_2$, the shift of Raman peak could also be caused by thickness difference of $MoS_2$ [17] at examined locations. Using the relationship between Raman peak shift and the number of (relaxed) $MoS_2$ layers reported in [17] and that of strain-induced Raman shift in $MoS_2$ reported in [22], we can uniquely determine the number of layers and the strain from the amount of redshift of $E_{2g}$ and $A_{1g}$ peaks observed in **Figure 5c** by solving two unknowns (i.e. strain and number of layers) from two equations describing the shift of $E_{2g}$ and $A_{1g}$ peaks, respectively. In particular, the in-plane mode $E_{2g}$ is more affected by the biaxial strain than out-of-plane mode $A_{1g}$ [22]. This would lead to a larger separation of the two peaks under biaxial tensile strain. The detailed calculation is discussed in the Supporting Information Section 5. With these analyses, we found that the Raman data of both 115 nm and 174 nm $SnO_x$/$MoS_2$/quartz samples in **Figure 5c** are from monolayer $MoS_2$. For the 115 nm $SnO_x$/$MoS_2$/quartz sample, 1.7% biaxial tensile strain is introduced by the $SnO_x$ stressor layer, which is expected to decrease the band gap by 0.17 eV [22]. For 174 nm $SnO_x$/$MoS_2$/quartz, 3.5% biaxial tensile strain is induced which would lead to 0.35 eV band gap shrinkage [22]. These data are largely consistent with the absorption peak redshift by ~0.2 eV and ~0.5 eV in these two samples, respectively, compared to pristine $MoS_2$ (see **Figure 5b**). The strain-induced band gap shrinkage especially enables absorption enhancement in NIR regime. In this way, the band gap tuning of $MoS_2$ avoids complicated device structure or external setup such as mechanical bending [23]. Furthermore, we achieve both photon management and strain engineering using a conductive electrode stressor layer of $SnO_x$ nanoneedle structure, which serves triple functions simultaneously and can be directly integrated with semiconducting 2D $MoS_2$ for device applications.

**Conclusion**

In summary, we report a synergistic approach for photon management and band gap engineering of 2D $MoS_2$. This approach utilizes the ultrahigh refractive index of Sn from visible to NIR light regime for photon management, which is demonstrated by two semimetal composite nanostructures: pseudo-periodic Sn nanodot and $SnO_x$ (x<1) nanoneedles. Both nanostructures are self-assembled and compatible with large scale manufacturing. With either of them coated on 2D $MoS_2$, the region with $MoS_2$ shows a clearly darker color contrast than that of its surrounding part without $MoS_2$ under white light illumination. The enhanced absorption has been quantitatively measured by spectrophotometer and verified by PL or Raman peak intensity enhancement. 2D $MoS_2$ shows up to 15x enhancement in absorption at λ=650-950 nm under pseudo periodic Sn nanodots and 20-30x at λ=700-900 nm under $SnO_x$ (x<1) nanoneedle. The $SnO_x$ (x<1) nanoneedle structure also serves as a conductive electrode and introduces tensile strain to $MoS_2$ at the same time, which decreases the band gap to further extend light absorption into the NIR regime. Therefore, $SnO_x$ nanoneedles uniquely achieve a *triple functional* feature for photon management, strain-induced band engineering, and conductive electrode layer on 2D $MoS_2$ for the first time. Such synergistic photon management and band gap engineering approach for extended spectral response can be further applied to other 2D materials for future 2D photonic devices. The investigation of semimetal photonic structures also paves the way of developing photonic functionalities beyond high index dielectrics and plasmonics.

## Methods

**Preparation of Semimetal Composite Nanostructures**

Semimetal composite nanostructures investigated in this paper are self-assembled from physical materials processing. Pseudo-periodic Sn nanodots are prepared by thermal evaporation process, while $SnO_x$ (x<1) nanoneedles are prepared by co-sputtering of Sn and $SnO_2$ in a ratio of 11:2 followed by low temperature anti-oxidation annealing process. Details of the processing methods have been previously reported in Ref. 11-13 and Ref. 19-20.

**Preparation of 2D $MoS_2$ on Various Substrates**

Chemical vapor deposition is used to make $MoS_2$ samples (CVD). The sulfur precursor (15 mg sulfur powder in an alumina boat) is placed on the furnace's upstream side. In the furnace center, a molybdenum precursor (10 mg $MoO_3$, Alfa Aesar, 99.9%) is inserted in an alumina boat. On the $MoO_3$ boat, $SiO_2$/Si substrates were mounted facedown. Following a purge with 1000 sccm Ar, the furnace was heated to 625 °C in 15 minutes under 20 sccm Ar. The temperature was held for 3 min to enable $MoS_2$ growth, following the immediate opening of the furnace for rapid cooling.

**Optical Characterization**

The transmittance and reflectance spectra are measured by a Jasco V-570 spectrometer equipped with a Jasco ISN-470 integrating sphere, scanned from λ=300 nm to 2500 nm.

**Material Surface Morphology Characterization**

A Veeco/Digital Instruments Dimension 3100 AFM is used for morphological characterization.

**Cross-Sectional Sample Preparation and Characterization**

A FEI Scios2 SEM-FIB dual beam system is used in cross-sectional sample preparation. A FEI Tecnai F20ST field emission gun system is used to capture the TEM image of the cross-sectional sample.

## Supporting Information

Method of deriving $MoS_2$ and $SnO_x$ refractive index from the experimental data, transmittance spectrum of the Sn nanodot/$MoS_2$/quartz sample, HRTEM of the $SnO_x$ nanoneedle structures, and $MoS_2$ strain analyses are included in the Supporting Information.

## Acknowledgement

This work has been sponsored by National Science Foundation under the collaborative research awards #1509272 and #1509197. We greatly appreciate the advanced characterization instruments of the Electron Microscope Facility at Dartmouth College and the Center of Nanoscale Systems in Harvard University. We also appreciate the help from Dr. Austin Akey for FIB machine operation.

**Table of Contents**

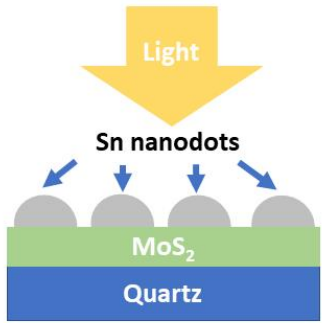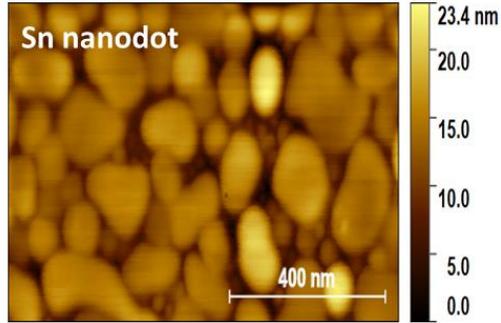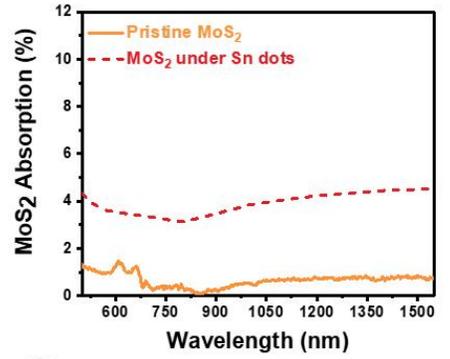
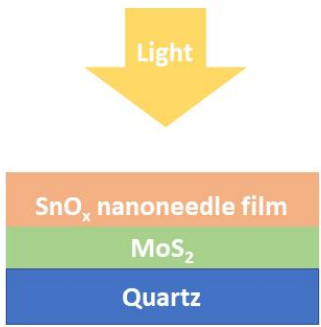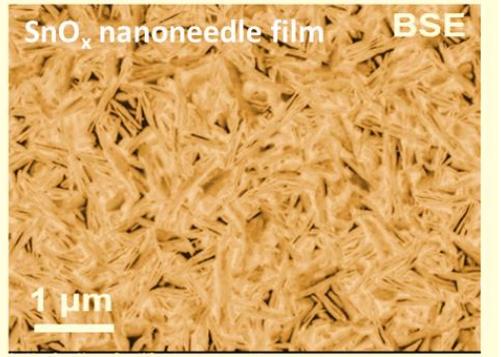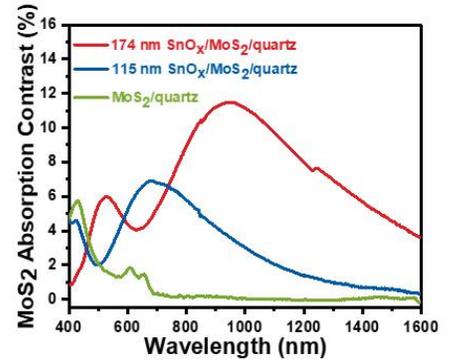

Supporting Information for

# Synergistic Photon Management and Strain-Induced Band Gap Engineering of Two-Dimensional MoS$_2$ Using Semimetal Composite Nanostructures

Xiaoxue Gao[1,*], Sidan Fu[1,*], Tao Fang[1], Xiaobai Yu[1], Haozhe Wang[2], Qingqing Ji[2,3], Jing Kong,[2,**] Xiaoxin Wang,[1,**] Jifeng Liu[1,**]

[1] Thayer school of Engineering, Dartmouth College, 15 Thayer Drive, Hanover, New Hampshire 03755, USA

[2] Department of Electrical Engineering and Computer Science, Massachusetts Institute of Technology, 77 Massachusetts Avenue, Cambridge, Massachusetts 02139, USA

[3] School of Physical Science and Technology, ShanghaiTech University, Shanghai 201210, China (current address)

[*] Authors contribute equally

[**]Corresponding authors:    Jifeng.Liu@dartmouth.edu;    Xiaoxin.Wang@dartmouth.edu; JingKong@mit.edu

## 1. Refractive Index of MoS$_2$

The MoS$_2$ refractive indexes n and k used in the simulation were derived by the optical transition matrix model and experimental measured absorption, reflection, and transmission data for MoS$_2$ on the SiO$_2$ substrate. In the measurement, the incident light is perpendicular to the MoS$_2$ layer. The absorption by the MoS$_2$ layer is

$$A = 1 - e^{\alpha l}$$

Here l is the thickness of the MoS$_2$ layer and α is the absorption coefficient. Since k is almost zero for MoS$_2$, the absorption coefficient is proportional to the imaginary refractive index k,

$$\alpha = \frac{4\pi}{\lambda} k$$

Here $\lambda$ is the wavelength of the incident light.

Therefore,

$$A = 1 - e^{\frac{4\pi}{\lambda} kl} \approx \frac{4\pi kl}{\lambda}$$

The equation is simplified because the thickness of the MoS$_2$ layer is much smaller than the wavelength of the incident light.

After obtaining the k values, the real part n can be derived by the optical transfer matrix and transmission data,

$$T = \frac{\beta'}{\beta} |t_1|^2 |t_2|^2 e^{-\alpha l}$$

Here, $\beta'$ and $\beta$ are the wavevectors in the final medium and initial medium separately. For this system, this ratio is 1.46. The $t_1$ is the transmission factor in the air-MoS$_2$ interface and $t_2$ is the transmission factor in the MoS$_2$-SiO$_2$ interface,

$$t_1 = \frac{2}{1 + n + ik}$$

$$t_2 = \frac{2(n + ik)}{n + ik + 1.46}$$

Using the k values derived previously from the absorption spectrum, we can solve the n value by this equation.

One restriction for this calculation is that only the first order absorption and transmission is considered. The multi-reflection effect is omitted in this calculation. However, for the MoS$_2$ system, the higher-order effect is much smaller than the first order effect, because the n and k values are not large. Comparing to the values in the literature, our n value is almost the same and k value is slightly smaller than that [1]. Therefore, this approximation is valid.

## 2. Transmittance spectrum of the Sn nanodot/MoS$_2$/quartz sample

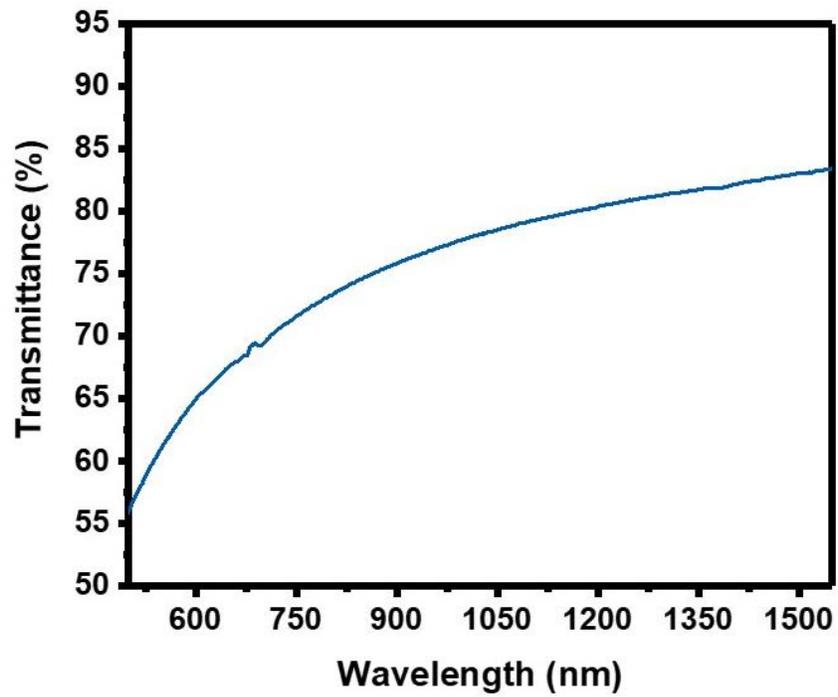

**Figure S1.** transmittance spectrum of Sn nanodot/MoS$_2$/quartz sample

## 3. TEM of SnO$_x$ nanoneedle structure

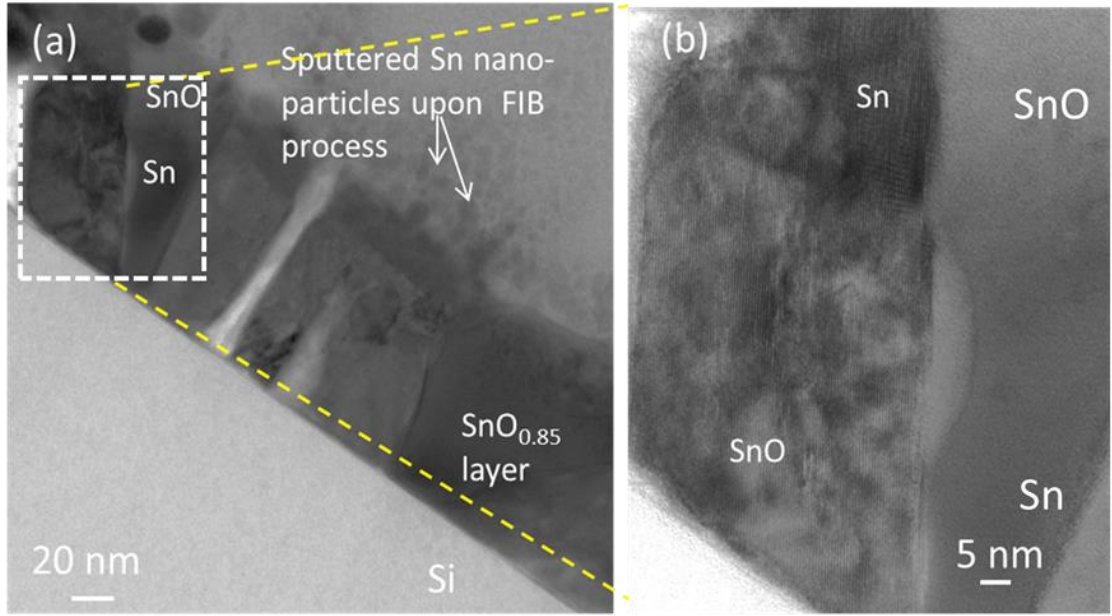

**Figure S2.** (**a**) TEM photo of SnO (x<1) nanoneedle cross section. Rectangular/Diamond shaped cross-section of nanoneedle is observed. Core-shell structure can be clearly observed in one of the nanoneedle cross-section, as indicated by white dash square. (**b**) High resolution TEM image of the region marked by yellow dashed line in (**a**). The dark color region is Sn phase while the light color region is SnO phase.

## 4. Refractive index of the SnO$_x$ thin film

Assuming the SnOx thin film is uniform, we could try to understand the photon management effect of this thin film by the derivation of its refractive index. The refractive index of SnOx thin film changes with its thickness which could be explained by the stoichiometry difference between SnOx thin film with different thickness. As reported previously by our research group, the value x in SnOx for 115 nm and 174 nm thin film is 1.02 and 0.75 respectively [2]. Thus, compared with 115 nm SnOx thin film, the 174 nm one contains more Sn in the film.

The blue and red curves in **Figure S3a** and **Figure S3b** are measured transmittance and reflectance spectra of 115 nm and 174 nm SnOx thin films on fused quartz substrate respectively. From these experimentally measured spectra, the refractive index, n and k, of SnOx thin film could be derived by combining optical transfer matrix and Sellmeier equation [3]. Here, we assume the thickness of fused quartz to be infinite since it is orders of magnitude larger than that of SnOx film. First of all, the refractive index at λ<650 nm and λ>1400 could be derived by optical transfer matrix. The refractive index between 650 nm and 1400 nm could not be solved by optical transfer matrix because it has no numerical solution in this range. Then these computed refractive indices are applied in the following Sellmeier equation to calculate the Sellmeier coefficients, $A_1$, $B_1$, $C_1$, $D_1$, $E_1$:

$$n^2(\lambda) = A_1 + \frac{B_1 \lambda^2}{\lambda^2 - C_1} + \frac{D_1 \lambda^2}{\lambda^2 - E_1}.$$

where n is the real part of the refractive index, λ is the wavelength of incident light in vacuum. With derived Sellmeier coefficients, the refractive index between 650 nm and 1400 nm could also be deduced.

**Figure S3c** demonstrates the derived refractive index of SnOx thin film. For 115 nm thick SnOx thin film, n decreases in visible light regime and keeps relatively flat in near IR light regime. For 174 nm thick SnOx thin film, n starts to increase at around 700 nm. This is consistent with our previous observation that Sn phase gets richer in the film as the SnOx film becomes thicker. In other words, Sn would contribute more significantly to the overall refractive index in thicker film.

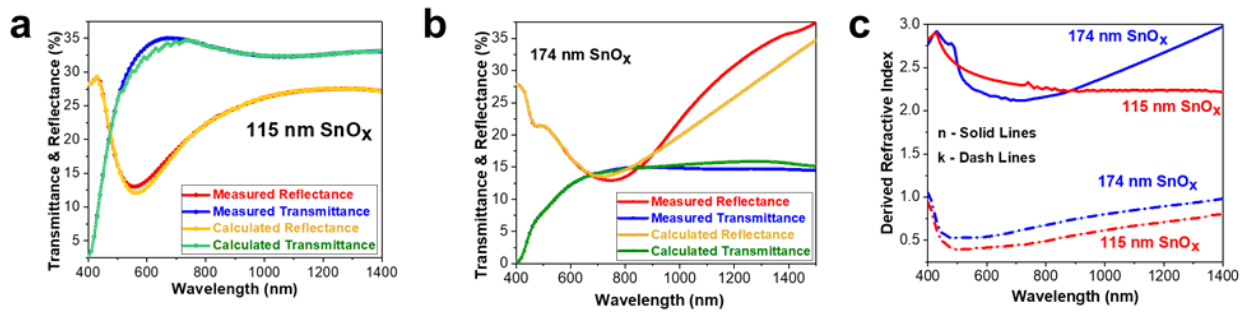

**Figure S3 (a-b)** Blue and red curves: experimentally measured transmittance and reflectance spectra of $SnO_x$ (x<1) thin films with the thickness of **(a)** 115 nm and **(b)** 174 nm respectively. Green and yellow curves: recalculated transmittance and reflectance spectra of $SnO_x$ (x<1) thin films with the thickness of **(a)** 115 nm and **(b)** 174 nm respectively. **(c)** Derived refractive indices of $SnO_x$ (x<1) thin films of thickness 115 nm and 174 nm from the experimentally measured transmittance and reflectance spectra, using optical transfer matrix method and Sellmeier equation.

With the derived refractive index from visible to near IR light regimes in **Figure S3c**, we can recalculate the transmittance and reflectance spectra through optical transfer matrix. These calculated spectra are shown by the green and yellow curves in **Figures S3a and Figure S3b**. For 115 nm thick $SnO_x$ film, the recalculated spectra match the measured spectra very well. However, for 174 nm thick $SnO_x$ thin films, there is some discrepancy between the measured and calculated spectra. It indicates that the scattering effect of Sn nanostructure becomes significant for 174 nm thick $SnO_x$ film. The assumption of treating $SnO_x$ layer as a uniform film would not be appropriate for thick $SnO_x$ film. Models considering scattering effect would be needed to understand the photon management of thick $SnO_x$ film.

## 5. Calculation of strain introduced in 2D $MoS_2$ by $SnO_x$ nanoneedle structure

The two most dominant Raman peak of single layer, unstrained $MoS_2$ is one in-plane $E_{2g}$ peak at 385 $cm^{-1}$ and one out-of-plane $A_{1g}$ peak at 405 $cm^{-1}$. In **Figure 5c**, large red shift of $E_{2g}$ and $A_{1g}$ Raman peak are observed. However, both the thickness of 2D $MoS_2$ and strain introduced in 2D $MoS_2$ could cause shift of Raman peak position [4-5]. To quantitatively analyze the strain introduced into the $MoS_2$, the relationship between the shift of Raman peak position, thickness of $MoS_2$ and strain is deduced first. As shown in the experimental work of Lloyd, David, et al. [5], the relation between shift of Raman peak position and strain is linear. In this linear relationship, the slope and y intercept are the function of thickness so we can construct the following equation to represent the relation:

$$f(x, y) = f_1(y) * x + f_2(y)$$

where $f(x, y)$ is the shift of Raman peak position, x is strain induced in $MoS_2$ and y is the number of $MoS_2$ layer. $f_1(y)$ is the relationship between Raman peak and strain which could be fit from the data in Ref. 5 and Ref. 6. $f_2(y)$ represents the relationship between Raman peak and thickness without any strain which could be fit from the data in Ref. 4.

Since we have two characteristic Raman peak positions in MoS2 system, we could construct two equations to solve for the two unknown variables x and y from experimental data. For example, the two equations to solve for x and y for 115 nm $SnO_x$ on $MoS_2$ sample are shown below:

$$E2g: f(x, y) = (0.366y^2 - 0.481y - 5.06) * x + \frac{2.57}{y} - 2.52$$

$$A1g: g(x, y) = (0.528y^2 - 1.57y - 0.745) * x - \frac{5}{y} + 4.78$$

where $f(x, y)$ here is the shift of $E2_g$ peak position and $g(x, y)$ represents the shift of $A1_g$ peak position. By this model, the number of $MoS_2$ layer and the induced strain could be calculated from the shift of the Raman peak position.

The introduced strain is classified into two types: compressive and tensile. And each type could be further classified into two kinds of directions: uniaxial and biaxial. Both compressive and tensile strain could cause shift of Raman peak position, but the trend of shift is different. Tensile strain leads to downshift of Raman peak position while compressive strain causes upshift of Raman peak position [7]. According to the trend shown in **Figure 5c**, the strain introduced here by $SnO_x$ nanoneedle is tensile. To analyze the direction of this tensile strain, we apply the mathematical model as stated in the last paragraph to fit both the uniaxial and biaxial tensile strain data in Ref.

5 and Ref. 6. The uniaxial tensile strain fitting shows that the number of $MoS_2$ layer is 0.4 which is not a reasonable result. Therefore, we accept the biaxial tensile strain fitting result: for 115 nm $SnO_x$ on $MoS_2$ sample, the area characterized by Raman spectroscopy is single layer $MoS_2$ induced with 1.7% biaxial tensile strain. As reported in the experimental work of Ref. 5, the 1.7% biaxial tensile strain could decrease optical band gap of monolayer $MoS_2$ at this area by 0.169 eV. Similarly, for 174 nm $SnO_x$ on $MoS_2$ sample, the characterized area is also single layer $MoS_2$ induced with 3.5% biaxial tensile strain. The optical band gap is lowered by 0.346 eV in this case.